\documentclass[aip]{revtex4-1}
\usepackage[utf8x]{inputenc}
\usepackage{amsmath}
\usepackage{amssymb}
\usepackage{amsfonts}

\begin{document}

\title{Discrete-time quantum walks: continuous limit and symmetries}
\author{G. di Molfetta}
\affiliation{Universit\'e Paris 6, ERGA-LERMA, UMR 8112, 3, rue Galil\'ee, F-94200 Ivry, France}
\author{F. Debbasch}
\email{fabrice.debbasch@gmail.com}
\affiliation{Universit\'e Paris 6, ERGA-LERMA, UMR 8112, 3, rue Galil\'ee, F-94200 Ivry, France}

\begin{abstract}
The continuous limit of one dimensional discrete-time quantum walks with time-and space-dependent coefficients is investigated. A given quantum walk does not generally admit a continuous limit but some families (1-jets) of quantum walks do.
All families (1-jets) admitting a continuous limit are identified.
The continuous limit is described by a Dirac-like equation or, alternately, a couple of Klein-Gordon equations. Variational principles leading to these equations are also discussed, together with local invariance properties.
\end{abstract}

\pacs{03.65.-w Quantum Mechanics, 03.67.-a Quantum information, 
5.60.-k Transport proceses}
\keywords{Random walks, Dirac equation, Klein-Gordon equation}

\maketitle

\section{Introduction}

Quantum walks are the simplest formal analogues of classical random walks. 
They have been introduced in their discrete-time version by \cite{ADZ93a} and \cite{Meyer96a} and the continuous-time version first appeared in \cite{FG98a}.
 Quantum walks are important in quantum information and quantum computing, where several algorithms are based on them; examples include 
Ambainis's algorithm for element distinctness \cite{Amb07a} and an algorithm for the triangle finding problem \cite{MNRS07a}. Quantum walks are also useful in several physical or biological contexts. They are naturally systems of choice to explore fundamental issues in quantum physics, including decoherence \cite{var96a,Perets08a}, but they also prove useful in modeling transport in solids \cite{Bose03a,Aslangul05a,Burg06a, Bose07a}, disordered media \cite{AVG98a,Mourach98a,Wester06a} and even complexes of algae \cite{Engel07a,Collini10a}. Quantum walks have been realized experimentally, for example as transport of  trapped ions\cite{Schmitz09a,Zahring10a}, of photons in wave guide lattices \cite{Perets08a} or optical networks\cite{Schreiber10a} and of atoms in optical lattices \cite{Karski09a}. Quantum walk experiments of two photons \cite{Peruzzo10a} have recently been performed, with the possibility of simulating Bose or Ferni statistics \cite{Sansoni11a}.
Cavity QED quantum walks have also been proposed \cite{Sanders03a}.

The continuous limit of discrete time quantum walks with constant coefficients has already been addressed by several authors \cite{FeynHibbs65a,KRS03a,BH04a,Strauch06a,Chandra10a}, with very different and apparently not reconcilable approaches. This article revisits this problem and obtains new, definitive conclusions for one-dimensional discrete-time quantum walks, which are entirely  
determined by three possibly time- and space-dependent Euler angles \cite{Chandra10a}. 
Our main results are the following: (i) generically, the object which can admit a continuous limit is {\sl not} a given discrete-time walk, but rather a 1-jet of discrete-time walks, which regroups all walks which share a common expansion in the time-and length steps as these tend to zero (ii) only a few 1-jets admit a continuous limit (iii) when it exists, the continuous limit can be described by a $2D$ Dirac -like equation or, alternately, two independent Klein-Gordon equations (iv) the Dirac-like equation derives from a variational principle which exhibits an interesting symmetry.

\section{Fundamentals}

\subsection{Discrete-time quantum walks}

We consider quantum walks defined over discrete time and discrete one dimensional space, driven by
time- and space-dependent quantum coins acting on a two-dimensional Hilbert space $\mathcal H$. The walks are defined by the following finite difference equations, valid for all $(j, m) \in \mathbb{N}  \times \mathbb{Z}$: 
\begin{equation}
\begin{bmatrix} \psi^{-}_{j+1, m }\\ \psi^{+}_{j+1, m } \end{bmatrix} \  = 
B\left( \theta_{j, m} ,\xi_{j, m} ,\zeta_{j, m} \right)
 \begin{bmatrix} \psi^{-}_{j, m+1} \\ \psi^{+}_{j, m-1} \end{bmatrix},
\label{eq:defwalkdiscr}
\end{equation}
where 
\begin{equation}
 B(\theta ,\xi ,\zeta) = 
\begin{bmatrix}  e^{i\xi} \cos\theta &  e^{i\zeta} \sin\theta\\ - e^{-i\zeta} \sin\theta &  e^{-i\xi} \cos\theta
 \end{bmatrix}
\label{eq:defB}
\end{equation} is a $\mbox{SU}(2)$ operator defined in terms of its three Euler angles $\theta, \xi, \zeta$.
The index $j$ labels instants  and the index $m$ labels spatial points.
The wave function $\Psi$ has two components $\psi^- $ and $\psi^+$ which code for the probability amplitudes of the particle jumping towards the left or towards the right.
The total probability $\pi_j= \sum_m \left(
\mid \psi^- _{j, m} \mid ^2+ \mid \psi^+ _{j, m}\mid^2 \right)$ is independent of $j$ {\sl i.e.} conserved by the walk.
The set of angles $\left\{ \theta_{j, m} ,\xi_{j, m} ,\zeta_{j, m}, (j, m) \in \mathbb{N}  \times \mathbb{Z}\right\}$ defines the walk and is, at this stage, arbitrary.

To investigate the continuous limit,
we first introduce a time step $\Delta t$ and a space step $\Delta x$. We then introduce, for any quantity $a$ appearing in (\ref{eq:defwalkdiscr}), a function $\tilde a$ defined on $\mathbb{R}^+ \times \mathbb{R}$ such that the number $a_{j,m}$ is the value taken by $\tilde a$ at the space-time point $(t_j = j \Delta t, x_m = m \Delta x)$. Equation (\ref{eq:defwalkdiscr}) then reads:
\begin{equation}
\begin{bmatrix} \psi^{-}(t_j+\Delta t, x_m) \\ \psi^{+}(t_j+\Delta t, x_m) \end{bmatrix} \  = 
B\left(\theta(t_j, x_m) ,\xi(t_j, x_m) ,\zeta(t_j, x_m) \right)
 \begin{bmatrix} \psi^{-}(t_j, x_m+\Delta x) \\ \psi^{+}(t_j, x_m-\Delta x) \end{bmatrix},
\label{eq:defwalk}
\end{equation}
where the tilde's have been dropped on all functions to simplify the notation. 
We now suppose, that all functions can be chosen at least $C^2$ in both space and time variables for all sufficiently small values of $\Delta t$ and $\Delta x$.  The formal continuous limit is defined as the couple of differential equations obtained from ({eq:defwalk})  by letting both $\Delta t$ and $\Delta x$ tend to zero.

\subsection{Scaling for the continuous limit}

Let us now introduce a time-scale $\tau$, a length-scale $\lambda$, an infinitesimal $\epsilon$ and write
\begin{eqnarray}
\Delta t & = & \tau \epsilon \nonumber \\
\Delta x & = & \lambda \epsilon^\delta,
\end{eqnarray}
where $\delta >0$ traces the fact that $\Delta t$ and $\Delta x$ may tend to zero differently.

For the continuous limit to exist, at least formally, the operator $B({\theta ,\xi ,\zeta})$ defined by the functions $\theta$, $\xi$ and $\zeta$ must also tend to unity as $\epsilon$ tends to zero. This is so because the two column vectors of the left-hand side and on the right-hand side of (\ref{eq:defwalk}) both tend to $\Psi(t_j, x_m)$  when $\Delta t$ and $\Delta x$ tend to zero. This  trivial remark implies that
the two functions $\theta$ and $\xi$ must 
actually depend on $\epsilon$ and 
tend to 0 or $\pi$ as $\epsilon$ tends to zero. In other words, it generally does not make sense to consider the continuous limit of a given walk, defined by $\epsilon$-independent angles, but rather the limit of a family of walks indexed by $\epsilon$, whose defining angles depend on $\epsilon$ and tend to zero with this infinitesimal.
Since we are only interested in the behavior of the family as $\epsilon$ tends to zero, we only need to consider families which are jets and we thus write:
\begin{eqnarray}
\theta(t, x)& = & p \pi + {\bar \theta}(t, x) \epsilon^\alpha \nonumber \\
\xi(t, x)& = & p\pi + {\bar \xi}(t, x) \epsilon^\beta,
\end{eqnarray}
where $p = 0$ or $1$ and $\alpha, \beta >0$. 
Note that the third Euler angle $\zeta$ does not have to tend to zero with $\epsilon$ because of the $\sin \theta$ factor in front of $e^{i \zeta}$ in (\ref{eq:defB}). Note also that the Hadamard walk cannot have a continuous limit. This point is further discussed in Section 5 below.

The continuous limit can then be investigated by Taylor expanding $\psi^\pm (t, x  \mp \Delta x)$, 
$\psi^\pm (t \pm \Delta t, x)$, $\cos \theta$, $\sin \theta$ and $e^{i \xi}$ in powers of $\epsilon$.  
The possible scalings obeyed by the continuous limit are found by examining the lowest order contributions.
One has:
\begin{equation}
\psi^\pm (t \pm \Delta t, x) = \psi^\pm (t, x) + O(\epsilon),
\end{equation}
\begin{equation}
\psi^\pm (t, x \mp \Delta x) = \psi^\pm (t, x) + O(\epsilon^\delta),
\end{equation}
\begin{equation}
e^{i\xi} \cos\theta = 1 + O(\epsilon^\beta) + O(\epsilon^{2 \alpha}),
\end{equation}
and
\begin{equation}
e^{i\zeta} \sin\theta =  O(\epsilon^\alpha).
\end{equation}
Equation (\ref{eq:defwalk}) then leads to:
\begin{equation}
\begin{bmatrix} \psi^{-}(t_j, x_m) \\ \psi^{+}(t_j, x_m) \end{bmatrix} \ + O(\epsilon) = 
\begin{bmatrix} \psi^{-}(t_j, x_m) \\ \psi^{+}(t_j, x_m) \end{bmatrix} + O(\epsilon^\alpha) + O(\epsilon^{\beta}) +  O(\epsilon^\delta).
\label{eq:basescal}
\end{equation}

Zeroth order contributions cancel out as expected and the remaining terms must balance each other. 
The richest and most interesting case corresponds to $\alpha = \beta = \delta = 1$ because all contributions 
to (\ref{eq:basescal}) are then of equal importance. This is the scaling that will be investigated in the remaining of this article.

\section{Equations of motion}

The natural space-time coordinates to investigate this scaling are the so-called null coordinates $u^-$ and $u^+$, defined in terms of $x$ and $t$ by  
\begin{eqnarray}
u^- & = & \frac{1}{2} \left(\frac{t}{\tau} - \frac{x}{\lambda} \right) \nonumber \\
u^+ & = & \frac{1}{2} \left(\frac{t}{\tau} + \frac{x}{\lambda} \right).
\end{eqnarray}
Partial derivatives with respect to these null coordinates read:
\begin{eqnarray}
\partial_- & = & \partial_{u^-}  = \tau \partial_t -  \lambda \partial_x   \nonumber \\
\partial_+ & = & \partial_{u^+} =   \tau \partial_t+ \lambda \partial_x,
\label{eq:partialuv}
\end{eqnarray}
and 
\begin{equation}
\partial_-\partial_+  =  \tau^2 \partial_{tt} - \lambda^2 \partial_{xx} =  \lambda^2 \Box,
\end{equation}
where $\Box $ is the usual d'Alembert operator, defined by
\begin{equation}
\Box = \frac{1}{c^2}\partial_{tt} -  \partial_{xx},
\end{equation}
with $c = \lambda/\tau$.

The expansion of the discrete equations around $\epsilon = 0$ then leads to the following
equations of motion for $\psi^-$ and $\psi^+$\, :
\begin{equation}
 \partial_-\psi^-  = \  \  \ \left( i  \overline{\xi}\psi^- - (-1)^{p+1} \overline{\theta} e^{+i\zeta}\psi^+ \right)
\label{eq:Dirac-}
\end{equation}
and 
\begin{equation}
 \partial_+\psi^+  =   - \left( i  \overline{\xi}\psi^+ - (-1)^{p+1} \overline{\theta} e^{-  i\zeta}\psi^- \right). 
\label{eq:Dirac+}
\end{equation}
These coupled first-order equations are best transcribed in operator form {\sl i.e.} as $\mathcal D \Psi = 0$,
with the operator $\mathcal D$ acting on the two-component wave function $\Psi$ given by:
\begin{equation}
\mathcal D = \Gamma^- \partial_- + \Gamma^+ \partial_+
- i \sigma_3 {\bar \xi} 
 + i \sigma_2
(-1)^{p+1} {\bar \theta} \cos \zeta  + i \sigma_1 (-1)^{p+1} {\bar \theta} \sin \zeta,
\label{eq:defD}
\end{equation}
where the $\sigma$'s are the three Pauli matrices:
\begin{equation}
\sigma_1 = 
\begin{bmatrix}  0 &  1\\ 1 &  0
 \end{bmatrix}
\label{eq:defsigma1}
\end{equation}
\begin{equation}
\sigma_2 = 
\begin{bmatrix}  0 &  -i\\ i &  0
 \end{bmatrix}
\label{eq:defsigma2},
\end{equation}
\begin{equation}
\sigma_3 = 
\begin{bmatrix}  1 &  0\\ 0 &  -1
 \end{bmatrix}.
\label{eq:defsigma3}
\end{equation}
and
\begin{equation}
\Gamma^\mp = \frac{1}{2}\ \left( 1 \pm \sigma_3\right).
\label{eq:defGamma}
\end{equation}
The operator $\mathcal D$ defines the formal continuous limit of the quantum walk. 
Equations (\ref{eq:Dirac-}) and  (\ref{eq:Dirac+}) are sometimes called the Dirac form of the continuous quantum walk dynamics.

These first-order equations imply that each component of $\Psi$ obeys, at points where $\theta$ does not vanish, an uncoupled Klein-Gordon equation. For example, the equation obeyed by $\psi^-$ can be obtained by using (\ref{eq:Dirac-}) to express $\psi^+$ in terms of $\psi^-$ and $\partial_- \psi^-$, and by then replacing 
$\psi^+$ by this expression in  (\ref{eq:Dirac+}). This leads to

\begin{eqnarray}
\lambda^2 \Box \psi^-&=& \left[ \partial_+ \left(\ln \bar \theta +i \zeta \right)-i \bar \xi\right] 
\partial_- \psi^- +i \bar \xi \partial_+\psi^- \nonumber \\ 
& & - \left[\bar \theta^2 +\bar \xi^2+i \bar \xi \partial_+\left(\ln \bar \theta+i \zeta \right)
-i\partial_+\bar \xi \right] \psi^- 
\label{eq:KG-})
\end{eqnarray}
and
\begin{eqnarray}
\lambda^2 \Box \psi^+&=&-i \bar \xi \partial_-\psi^++\left[ \partial_- \left(\ln \bar \theta -i \zeta \right)
+i \bar \xi\right] 
\partial_+ \psi^+  \nonumber \\ 
& & - \left[\bar \theta^2 +\bar \xi^2+i \bar \xi \partial_-\left(-\ln \bar \theta+i \zeta \right)
+i\partial_-\bar \xi \right] \psi^+ .
\label{eq:KG+})
\end{eqnarray}

These equations ressemble Klein-Gordon equations, but contain terms which violate time-reversibility. The discussion of these equations is postponed till section 4.2.

The natural initial conditions for the first order equations (\ref{eq:Dirac-}) and (\ref{eq:Dirac+}) are simply the values of $\psi^-$ and $\psi^+$ at time $t = 0$ {\sl i.e.} on the submanifold $u^- = - u^+$. The natural initial conditions for (\ref{eq:KG-}) are the values of $\psi^-$ and $\partial_t \psi^-$ on the sub-manifold $t = 0$. By equation (\ref{eq:Dirac-}), giving oneself these initial conditions is equivalent to giving oneself both $\psi^-$ and $\psi^+$ at time $t = 0$. The same reasoning applies to the initial conditions of (\ref{eq:KG+}).

\section{Variational principles and symmetries}

\subsection{First order form of the equations}

It is straightforward to check that the Lagrangian density:
\begin{equation}
L_D [\Psi, \Psi^\dagger] = \Psi^\dagger {\mathcal D} \Psi 
\label{eq:defLD}
\end{equation}
leads to the first-order equations of motion (\ref{eq:Dirac-}, \ref{eq:Dirac+}).
This density and, hence the dynamics, is {\sl not} invariant under the  (global or local) action of $SU(2)$. 
If it were, one would have $U \mathcal D U^{-1} = \mathcal D$ {\sl i.e.} $[\mathcal D, U] = 0$ for all elements $U$ of SU(2).
The generator $\sigma_3$ does commute with $\mathcal D$, but $\sigma_1$ and $\sigma_2$ do not. This reflects the fact that $\sigma_1$ and $\sigma_2$ interchange $\psi^-$ and $\psi^+$ while these two components of $\Psi$ play fundametally different roles, one propagating towards the left while the other propagates towards the right.
This in turn is reflected by the fact that $\Gamma^+ \ne \Gamma^-$.

The first-order dynamics can nevertheless be linked to interesting locally invariant operators. Let us consider the 
following family of differential operators:
\begin{equation}
\mathcal D (\mathcal B) = \Gamma^\mu \nabla_\mu (\mathcal B) = \Gamma ^\mu \left( \partial_\mu + i \sum_j \mathcal B^j_\mu \sigma_j \right)
\label{eq:defDB}
\end{equation}
where
$\mu \in \{ -, + \}$ is the space-time index, $j \in \{1, 2, 3\}$ and the $\mathcal B$'s are arbitrary real, possibly time- and position dependent quantities; the $\Gamma$ matrices have been defined above in (\ref{eq:defGamma}).

It is obvious that the operator $\mathcal D$ characterizing the continuous dynamics of the quantum walks considered in this article belongs to this family. Let us now consider, for any $j \in\{1, 2, 3\}$ and arbitrary time- and space-angle $\alpha$, the following transformation of $\mathcal D (\mathcal B)$:
\begin{equation}
{\mathcal D}(\mathcal B) _{j, \alpha} (\Psi) = \exp\left( - i \alpha \sigma_j\right) 
\left[
\mathcal D (\mathcal B) (\cos \alpha \Psi) 
+ i \sigma_j \mathcal D (\mathcal B) (\sin \alpha \Psi)
\right].
\label{eq:defDBjalpha}
\end{equation}
Replacing $\mathcal D (\mathcal B)$ by its definition (\ref{eq:defDB}) and using the fact that $\sigma_j^2 = 1$, one finds:
\begin{equation}
{\mathcal D}(\mathcal B)_{j, \alpha} (\Psi) = \mathcal D (\mathcal B)(\Psi) + 
i \sigma_j (\Gamma^\mu \partial_\mu \alpha) \Psi.
\end{equation}
In other words:
\begin{equation}
{\mathcal D}(\mathcal B)_{j, \alpha} = \mathcal D (\mathcal B({j, \alpha}))
\end{equation}
with 
\begin{equation}
\mathcal B_\mu^k ({j, \alpha}) = \mathcal B_\mu^k + \delta_j^k \partial_\mu \alpha. 
\label{eq:transconn}
\end{equation}
The family of operators $\mathcal D(\mathcal B)$ thus displays local invariance with respect to the transformation (\ref{eq:defDBjalpha}) and the operator fixing the continuous quantum 
walk dynamics is
one member of this locally invariant family. Let us stress that the operators $\mathcal D (\mathcal B)$ do {\sl not} commute with $\sigma_1$ and $\sigma_2$\footnote{because $\sigma_1$ and $\sigma_2$ interchange $\psi^-$ and $\psi^+$ while all operators $\mathcal D (\mathcal B)$ are built with $\Gamma^+$ and $\Gamma^-$ which are not identical; see the discussion above at the beginning of this section} and that the invariance considered here is not an invariance of the continuous quantum walk under the action of $SU(2)$. If it were, the $\sigma_j$ in the bracket on the right-hand side of equation (\ref{eq:defDBjalpha}) would appear on the right of  $\mathcal D (\mathcal B)$, not on the left. The invariance considered here is a formal invariance of an operator family, not an invariance of the equations of motions.  Note however that, for $j = 3$, the above transformation coincides with a standard gauge transformation along the $\sigma_3$ generator of $SU(2)$, because $\sigma_3$ commutes with all $\mathcal D (\mathcal B)$.

Let us remark that a generic $\mathcal D (\mathcal B)$ operator does not conserve the total probability
$\int_x dx [\mid \psi^- (t, x) \mid^2 + \mid \psi^- (t, x) \mid^2]$ for all $\psi^-$ and $\psi^+$. If one imposes total probability conservation as an identity valid for all values of $\psi^-$ and $\psi^+$, one obtains that 
$\mathcal D(\mathcal B)$ necessarily takes the form:
\begin{equation}
\mathcal D (\mathcal B) = \Gamma^\mu \partial_\mu - i \sigma_3 (\Gamma^\mu {\bar \xi}_\mu) 
 + i \sigma_2
(-1)^{p+1} {\bar \theta} \cos \zeta  + i \sigma_1 (-1)^{p+1} {\bar \theta} \sin \zeta  
\end{equation}
where $p$ is an arbitrary integer and $\xi_-$, $\xi_+$, $\bar \theta$ and $\zeta$ are arbitrary possibly time-and position dependent real quantities. This coincides with the Dirac form of the continuous quantum walk, except for the fact that $\xi_-$ is not necessarily identical to $\xi_+$. This can be remedied by exploiting the local $U(1)$ invariance of the problem {\sl i.e.} by changing the phase of $\Psi$. The operator $\mathcal D$ given by equation 
(\ref{eq:defD}) is thus the most general $\mathcal D (\mathcal B)$ operator which preserves the total probability as an identity of the dynamics. By (\ref{eq:transconn}) and (\ref{eq:defD}), this property is preserved under a transformation (\ref{eq:defDBjalpha}) if and only if $\partial_- \alpha = \partial_+ \alpha$ {\sl i.e.} $\partial_x \alpha = 0$.

Let us end this section by writing out two useful forms of the Lagrangian density $L_D$. The first form  serves as a reminder of the gauge invariance described above. It reads:
\begin{equation}
L_D[\Psi, \Psi^\dagger] = \Psi^* \Gamma^\mu \nabla_\mu \Psi
\end{equation}
with
\begin{equation}
\nabla_\mp = \partial_\mp - i \sigma_3 {\bar \xi} +  i \sigma_2
(-1)^{p+1} {\bar \theta} \cos \zeta  + i \sigma_1 (-1)^{p+1} {\bar \theta} \sin \zeta  
\end{equation}
This can be derived directly from (\ref{eq:defLD}) by noting that $\Gamma^- + \Gamma^+ = 1$.

The Lagrangian density $L_D$ can also be brought to a form which closely resembles the Dirac Lagrangian density. This is accomplished by singling out the ${\bar \xi} \sigma_3$ term, which then appears as a mass term (remember this term is related to the $U(1)$ invariance of $\Psi$). Let us define
\begin{equation}
\gamma_\mu  = - \sigma_3 \Gamma_\mu
\end{equation}
and 
\begin{equation}
D_\mu  = \partial_\mu + 
i \sigma_1 (-1)^{p+1} {\bar \theta} \sin \zeta  + 
i \sigma_2
(-1)^{p+1} {\bar \theta} \cos \zeta.  
\end{equation}
One can then write:
\begin{equation}
L_D[\Psi, \Psi^\dagger] = \Psi^\dagger \left(i \gamma^\mu D_\mu - {\bar \xi}\right)\Psi,
\end{equation}
which does look like the standard Lagrangian density for a Dirac field. Of course, $\Psi$ here belongs to a two -dimensional Hilbert space, and is not a Dirac spinor.

\subsection{Second order form of the equations}

Let us now investigate if each of the equations (\ref{eq:KG-}) and (\ref{eq:KG+}) can be obtained from a variational principle. As already noted, the second order equations (\ref{eq:KG-}) and (\ref{eq:KG+}) are not time-reversible. This irreversibility is made manifest in  (\ref{eq:KG-}) by the $(\partial_+ \ln {\bar \theta}) \partial_- \psi^-$ term and in (\ref{eq:KG+}) by the $(\partial_- \ln {\bar \theta}) \partial_+ \psi^+$ term.  We will only present autonomous variational principles for equations (\ref{eq:KG-}) and (\ref{eq:KG+}) in the special cases where these irreversible terms vanish. We nevertheless remark that a standard method for obtaining a variational principle for an irreversible dynamics is to enlarge the configuration space of the system and actually obtain a variational principle which delivers, not only the desired irreversible dynamics, but also another, coupled dynamics obeyed by some extra degrees of freedom. For the continuous quantum walks dynamics considered here, there are indeed no autonomous
variational principles which deliver separately the Klein-Gordon equations obeyed by $\psi^-$ and $\psi^+$ when disspative terms do not vanish, but there is always a single variational principle which delivers the coupled Dirac-like equations obeyd by the $\Psi = (\psi^-, \psi^+)$.

Let us start with equation (\ref{eq:KG-}) and suppose that  disspative terms vanish {\sl i.e.} that $\partial_+{\bar \theta} = 0$. The only reasonable candidate to build a variational principle for (\ref{eq:KG-}) is the standard Klein-Gordon Lagrangian density
\begin{equation}
L^-_{KG}[\psi^{- \dagger}, \psi^-] = g^{\mu \nu} (\nabla_\mu^- \psi^-)^\dagger \nabla_\nu^- \psi^- - M_-^2 
\mid \psi^- \mid^2
\label{eq:defLKG-}
\end{equation}
where
\begin{equation}
[g^{\mu \nu}] = 
\begin{bmatrix}  0 &  1\\ 1 &  0
 \end{bmatrix}
\end{equation}
are the metric components in the null coordinates $(u^-,u^+)$ and the covariant derivative $\nabla_\mu^- = \partial_\mu - i {\mathcal A}^-_\mu$ couples $\psi^-$ to a real potential $A^-$. Note also that the mass $M_-$ appearing in (\ref{eq:defLKG-}) must be real for the Hamilton equations obeyed by $\psi^-$ and $\psi^{- \dagger}$ to be equivalent, as they should.  The physical reason for this restriction on the mass $M_-$ is that a mass with imaginary part would generate an imaginary part in the frequency (energy) or wave-vector (impulse) of the particle, and this would signal irreversibility.

Identifying the equation of motion derived from (\ref{eq:defLKG-})  with equation (\ref{eq:KG-}) leads to:
\begin{eqnarray}
{\mathcal A}_-^-  & = & {\bar \xi} \nonumber \\
{\mathcal A}_+^-  & = & \partial_+ \zeta - {\bar \xi} \nonumber \\
M_- & = & \mid {\bar \theta} \mid
\end{eqnarray}
provided
\begin{equation}
\lambda^2 \Box \zeta = (\partial_- + \partial_+) {\bar \xi} = \tau \partial_t {\bar \xi}.
\label{eq:M-real}
\end{equation}
The only non possibly vanishing components of the electromagnetic tensor $F$ ({\sl i.e.} the curvature of the connection $\mathcal A$) are $F_{-+} = F_{+-}$. A straightforward calculation shows that these components actually vanish if the above conditions are satisfied. Thus, the potential $\mathcal A^-$ can be eliminated by the zero of phase for  $\psi^-$ and it does not represent a real, physical electromagnetic field. 

Proceeding in the same way for $\psi^+$ delivers 
\begin{eqnarray}
{\mathcal A}_-^+ & = &  - \left( \partial_- \zeta - {\bar \xi}\right) \nonumber \\
{\mathcal A}_+^+ & = & - {\bar \xi} \nonumber \\
M_+&  = & \mid {\bar \theta} \mid
\end{eqnarray}
under the conditions $\partial_- {\bar \theta} = 0$ and 
\begin{equation}
\lambda^2 \Box \zeta = - (\partial_- + \partial_+) {\bar \xi} = - \tau \partial_t {\bar \xi}.
\label{eq:M+real}
\end{equation}
As above, these conditions makes the electromagnetic tensor $F^+$ vanish. Thus, $\mathcal A^+$ does not represent a real, physical electromagnetic field.

We thus conclude that there are indeed situations for which one or the other autonomous Klein-Gordon equation follows from a variational principle, but that these situations are not particularly interesting, at least as far as the dynamics of  the retained component of $\Psi$ is concerned. In the other situations, each Klein-Gordon equation contains terms which induce time-irreversibility and the equations do not follow from autonomous variational principles. This is not so surprising after all. It only tells us that the two components of $\Psi$ are really coupled by the quantum walk and cannot be decoupled without paying the price {\sl i.e.} loosing symmetry and the possibility of writing simple variational principles.   

\section{Conclusion}

\subsection{Summary}
We have identified the 1-jets of  $1D$ discrete-time quantum walks which admit a continuous limit.
The continuous limit is described by a simple Dirac-like equation which derives from a variational principle.
Neither the dynamics of the discrete-time walk nor the dynamics of its continuous limit are invariant under $SU(2)$ , but the continuous dynamics nevertheless belongs to a family of Dirac-like equations which exhibit an elegant symmetry. In the continuous limit, each component of the wave function also obeys an autonomous Klein-Gordon equation, but these equations do not generally derive from variational principles.

\subsection{Discussion}

Let us now discuss how these results connect with the literature. As mentioned earlier, there are two types of quantum walks, called respectively discrete time \cite{ADZ93a,Meyer96a} and continuous time \cite{FG98a,Childs03a,MB11a} quantum walks. Space is discrete in both types of walks. It is shown in  \cite{Strauch06a} that, at least formally,
discrete time quantum walks become continuous time quantum walks when the time step tends to zero. The correspondence between discrete and continuous time quantum walks is also explored in \cite{Childs09a}. Our work investigates what happens when both time and length steps tend to zero. The problem was first touched upon by Feynman \cite{Schweber86a}, who was in search of a simple dynamics which, when quantized, would give back the Dirac equation.
Feynman's approach is discussed rapidly in \cite{FeynHibbs65a, Strauch06a} and the continuous limit of quantum walks is dealt with extensively in \cite{KRS03a,BH04a,Chandra10a}. 

References \cite{KRS03a,BH04a} propose continuous limits for several discrete walks with constant coefficients, including the Hadamard walk. This seems to contradict the material presented in Section 2 above, where we have proved by a rather trivial argument that the Hadamard walk cannot have a formal continuous limit. This apparent contradiction is resolved by
two remarks. First, the differential equations proposed in  \cite{KRS03a,BH04a} are not obeyed by the original field $\Psi$, but rather by two composite fields which mix $\psi^-$ and $\psi^+$ in a time dependent manner. Second,  the authors of references \cite{KRS03a,BH04a} do not use a strict concept of continuous limit. Indeed, the differential equations proposed in\cite{KRS03a,BH04a} are not the strict formal equivalent of the random walk equations when both time- and length steps tend to zero;  the differential equations proposed in \cite{KRS03a,BH04a} rather admit discretizations which start with terms identical to those found in the discrete quantum walk dynamics. This article therefore does not contradict references \cite{KRS03a,BH04a}, but rather complements them by presenting, when it exists, the strict formal continuous limit of discrete time-and space-dependent walks in the form of differential equations obeyed by $\Psi$.

References \cite{FeynHibbs65a, Strauch06a} address only the special case where $\theta$ and $\zeta$ identically vanish and $\xi$ is a constant. These references 
mention that a Dirac-like equation obeyed by $\Psi$ can be formally obtained from a discrete quantum walk when the angle $\xi$ tends to zero with the time and length steps, but they offer little detailed computations and do not mention the existence of Klein-Gordon equations.
Reference\cite{Chandra10a} only addresses situations where $\xi = \zeta = 0$ and $\theta$ is a constant. This reference does present a detailed computation and both Dirac and Klein-Gordon equations are obtained, but the crucial condition that $\theta$ must tend to zero together with the time and length step is overlooked.  As a result, the equations presented in \cite{Chandra10a} contain various trigonometric functions of $\theta$. Once the correct scaling law is applied, these equations turn out to be identical to the power laws and logarithm  appearing in our work.

The following four points are not discussed in the previous literature and have thus been addressed here for the first time: (i) general situations where none of the the Euler angle vanishes and where these angles depend on time and/or space (ii) the existence of several scaling laws (iii) the existence of variational principles for the continuous limit (iv) the symmetry properties of the continuous limit. The present article is thus a clear on the current literature.

This work can be extended in various directions. 
The above material makes clear that the scaling studied in this article is not the only one which allows for a continuous limit of the discrete walk. Other scalings, though not as rich as the one retained in this article, should naturally be studied in their own right. Detailed numerical simulations should also be performed to fully understand the dynamics of the continuous limits. It would also be very interesting to investigate the continuous limit of discrete quantum walks defined over a higher dimensional space-times and/or quantum walks driven by quantum coins acting on higher dimensional Hilbert spaces. And one can only wonders how discrete quantum walks can be coupled to space-time geometry and if this coupling can be of practical interest, for example in astrophysics or cosmology. It should finally prove very rewarding to investigate how decoherent continuous quantum walks connect with relativistic stochastic processes \cite{DC07b,DH08a}.

\def\cprime{$'$} \def\cprime{$'$} \def\cprime{$'$} \def\cprime{$'$}
  \def\cprime{$'$}

\end{document}